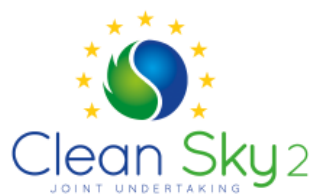 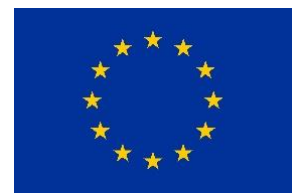

| **Publication title** | **Experimental and numerical study on the effect of oxymethylene ether-3 (OME$_3$) on soot particle formation** |
|---|---|
| **Authors** | Federica Ferraro, Carmela Russo, Robert Schmitz, Christian Hasse, Mariano Sirignano |
| **Issue Date** | 15 February 2021 |
| **Publisher** | Elsevier |
| **Type of publication** | Journal article |
| **Acknowledgement** | The ESTiMatE project has received funding from the Clean Sky 2 Joint Undertaking under the European Union's Horizon 2020 research and innovation programme under grant agreement No 821418. |
| **Disclaimer** | The content of this article reflects only the authors' view. The Clean Sky 2 Joint Undertaking is not responsible for any use that may be made of the information it contains. |

# Experimental and numerical study on the effect of oxymethylene ether-3 (OME$_3$) on soot particle formation

Federica Ferraro[1,*], Carmela Russo[2], Robert Schmitz[1], Christian Hasse[1], Mariano Sirignano[3]

1-Institute for Simulation of Reactive Thermo-Fluid Systems (STFS), Technische Universität Darmstadt, Otto-Berndt-Straße 2, Darmstadt 64287, Germany

2-Istituto di Ricerche sulla Combustione, Consiglio Nazionale delle Ricerche, P.le Tecchio 80, 80125 Napoli, Italy

3-Dipartimento di Ingegneria Chimica, dei Materiali e della Produzione Industriale – Università degli Studi di Napoli Federico II, P.le Tecchio 80, 80125 Napoli, Italy

**Abstract**
The reduction and control of particulate matter generated by fossil fuel combustion are among the main challenges in the design of future combustion devices. Recently different fuels have been investigated as a potential substitute or additive for diesel and gasoline. This work focuses on the effects of oxymethylene ether-3 (OME$_3$), the most promising of the OME compounds, on carbon particulate formation when blended with ethylene in burner-stabilized premixed flames at different equivalence ratios.
Particle size distribution (PSD) and Laser-Induced Fluorescence (LIF) and Incandescence (LII) along with numerical (Conditional Quadrature Method of Moments – CQMOM, based on D'Anna physico-chemical soot model) investigations were conducted to study particle formation and growth in pure ethylene and ethylene/OME$_3$ flames. Soot volume fraction and PSD indicate a reduction of the total amount and the size of the soot particles at all equivalence ratios, while the number of small nanoparticles remains almost unchanged. The CQMOM model is able to predict similar trends for the soot volume fraction and, using the entropy maximization concept, also the general shape of the PSD for both pure ethylene and OME$_3$ blended flames, compared to the experimental measurements.
Further, carbon particulate matter thermophoretically sampled in the highest equivalence ratio condition was spectroscopically analyzed. Soot structure was investigated using UV-Visible and Raman spectroscopy finding a slightly higher aromaticity for the pure ethylene soot. FT-IR analysis showed that carbon particulate matter produced from an OME$_3$-doped flame contained larger amounts of oxygen mainly in form of C=O.

Highlights:
- Soot particle formation in pure ethylene and ethylene/OME$_3$ premixed flames has been experimentally and numerally investigated.
- A reduction of the total amount and the size of soot particles is observed in OME$_3$ blended flames.
- Small nanoparticles (<10nm) are formed in similar numbers in both pure ethylene and ethylene/OME$_3$ flames.

* Corresponding author at: Institute for Simulation of Reactive Thermo-Fluid Systems (STFS), Technische Universität Darmstadt, Otto-Berndt-Straße 2, Darmstadt 64287, Germany.
E-mail address: ferraro@stfs.tu-darmstadt.de (Federica Ferraro)

- The aromaticity of carbon particulate is reduced when $OME_3$ is added.
- A higher presence of C=O functionalities was found on particles from $OME_3$ blended flames.

## 1. Introduction

Soot is a particulate pollutant generated by incomplete combustion of hydrocarbons. Due to its carcinogenic effects on human health and its detrimental impact on polar ice melting and climate change, the soot emission limits for combustion devices have become more stringent. Both soot mass concentration and particle size distribution need to be controlled.

While bigger particles can be filtered by exhaust gas after-treatment systems, small nanoparticles are more difficult to be trapped [1] and, once released in the atmosphere, they can penetrate deeper into the human respiratory system causing severe damage [2,3].

In order to reduce the carbon footprint as well as the particulate emissions of combustion systems, the use of alternative synthetic fuels has been largely explored [4,5]. Oxygenated fuels including molecules that can be mixed with diesel (DME, biodiesel, etc.) and gasoline (bioethanol, biobutanol, etc.) have been developed in the last years. The general belief is that due to their intrinsic structure the fuel is better oxidized [5] and less particulate pollutants are emitted [6,7]. Despite these advances, recent studies have shown that these fuels have indeed a great potential in reducing the total amount of particulate produced and hence emitted, however, particles are produced with smaller size [8] and, in some combustion conditions, in larger number concentration [4]. Additionally, recent findings have shown that the use of oxygenated fuels can lead to the production of oxy-PAH (polycyclic aromatic hydrocarbon) [9] and of particles with different chemical features, namely a larger presence of oxygen embedded onto particles [6,7,10]. This latter aspect could be responsible of the higher reactivity toward oxidation [6,8,11–16] of the particulate formed resulting in an easier abatement in the after-treatment systems but also in a higher propensity to interact with biological systems yielding a potential higher toxicity [3]. These controversial aspects have to lead the research of new alternative fuel candidates in order to prevent possible negative outcomes from their use on a large scale.

Among several alternative fuel candidates, oxymethylene ethers ($OME_n$), $CH_3O(CH_2O)_nCH_3$, also known as polyoxymethylene dimethyl ethers (PODE), are promising for realizing carbon-neutral combustion, used as additives or substitutes for diesel engines. OMEs belong to the class of e-fuels, i.e., fuels that can be produced by recycling $CO_2$ via electrolysis using renewable energies, contributing to the overall greenhouse gas balance [17]. The high oxygen content and the lack of C-C bonds of OMEs generate fewer soot precursors such as $C_2H_2$, $C_2H_4$ and $C_3H_3$, indicating a potential reduction of carbon particulates [18]. OMEs with $n > 1$ have high ignition propensity, with a cetane number exceeding 60 [18]. Additionally, OMEs are non-toxic and miscible in diesel fuel. OMEs have shown positive effects in CO, HC and soot particle emission reductions when used as an additive to diesel in engine experiments [19–25]. A higher rate of exhaust gas recirculation (EGR) can be applied to reduce $NO_x$ emissions [23,26]. Investigations on $OME_1$ as an alternative fuel indicate a total soot suppression [27]. However, since severe modifications are required for the engine and fuel supply system, blending OMEs in diesel fuels represents a more feasible strategy.

Previous studies have identified $OME_3$ as the smallest-sized $OME_n$ compound qualified for practical applications [18]. It has a lower melting and boiling point than $OME_4$ and a higher cetane number than $OME_1$ and $OME_2$. Furthermore, $OME_1$ is highly volatile and can vaporize in the storage, while $OME_2$ has a too low flash point [28].

Recently, reduced and detailed kinetic mechanisms have been developed for $OME_{1-3}$ [18,29,30] and for $OME_{2-4}$ [17], validated against experimental data for ignition delay time [17,18] and laminar flame speed [29]. The sooting propensity of OME/diesel blends in diffusion flames has been studied in [31].

Following previous studies on other alternative fuels [6,7,10,32–36], the primary goal of this work is the characterization of the sooting properties of $OME_3$ used as an additive in ethylene/air flames at different equivalence ratio [37]. In order to evaluate the effect of $OME_3$ and to identify some specific patterns related to fuel structure both experimental and numerical investigations are performed. To the best of our knowledge, there are no comprehensive studies on sooting behavior in laminar flames when $OME_3$ is used for blending other fuels. Experimental investigations include quantitative measurements of particulate both in situ (laser-based technique) and ex situ (particle size distribution measurements), and characterization of particulate (UV-Vis, Raman, FTIR analysis) in terms of composition and nanostructure. Simulations are performed with the detailed soot physico-chemical model proposed by D'Anna et al. [38,39] integrated into a conditional quadrature-based method of moments (CQMOM) approach [40,41]. Different types of species such as large PAHs, clusters and agglomerates are accounted for in the soot model, as well as dehydrogenation and oxidation-induced fragmentation processes. $OME_{1-3}$ kinetics from Sun et al. [29] has been included in the gas-phase kinetic mechanism to account for the fuel mixture oxidation. Classical CQMOM methods are not able to approximate the PSD directly and the Extended quadrature Method of Moments [42–44], has been recently introduced to cope with this problem. Following [41], an alternative approach based on the entropy maximization concept is employed here, which allows to reconstruct the PSD, giving a properly selected set of transported moments.

## 2. Experimental setup

Atmospheric pressure, premixed ethylene/air flames with equivalence ratio, ϕ, equal to 2.01, 2.16, 2.31 and 2.46 are stabilized on a capillary burner [37]. These flames are the reference cases for studying the effect of $OME_3$ blending on soot particle formation and growth. $OME_3$ was added by partially substituting the ethylene (20% of the total carbon fed) to the reference ethylene/air flames. The same experimental setup has been used for studying other alternative fuels as butanols [10], furans [33], ethanol [34], and dimethyl ether [36]. The combustion conditions investigated in this work are reported in Table 1.

Table 1 Flame conditions. Inflow mixture composition is given in mole fraction. Inlet Gas velocity @STP = 10 cm/s.

| ϕ | % $OME_3$ | 0 | 20 |
|---|---|---|---|
| 2.01 | $C_2H_4$ | 0.1234 | 0.0987 |
| | $O_2$ | 0.1841 | 0.1767 |
| | $N_2$ | 0.6925 | 0.7147 |
| | $OME_3$ | 0 | 0.0099 |
| 2.16 | $C_2H_4$ | 0.1313 | 0.1051 |
| | $O_2$ | 0.1824 | 0.1751 |
| | $N_2$ | 0.6862 | 0.7093 |
| | $OME_3$ | 0 | 0.0105 |
| 2.31 | $C_2H_4$ | 0.1392 | 0.1114 |
| | $O_2$ | 0.1808 | 0.1735 |
| | $N_2$ | 0.6800 | 0.704 |
| | $OME_3$ | 0 | 0.0111 |
| 2.46 | $C_2H_4$ | 0.1469 | 0.1175 |
| | $O_2$ | 0.1792 | 0.172 |
| | $N_2$ | 0.6739 | 0.6987 |
| | $OME_3$ | 0 | 0.0118 |

The experimental setup for optical and particle size distribution as well as the particulate collection on a glass plate is the same adopted in previous works (see [10,33,35]). Hereafter a brief description is reported for completeness. Laser-Induced Emission (LIE) measurements in the 200–550 nm range were used to detect particles in the flame, using as excitation source the fourth harmonic of a Nd:YAG laser at 266 nm [32,35–37]. The emitted spectra were collected with an ICCD camera with a gate of 100 ns that allow to distinguish between the broad Laser-Induced Fluorescence (LIF) signal, ranging between 300 and 450 nm, and the Laser-Induced Incandescence (LII) following a blackbody curve and evaluated at 550 nm.

In order to retrieve information on particle size distribution (PSD), particle sampling from the flames was performed with a horizontal probe [45–50]. The horizontal probe adopted here has an ID = 8 mm, a wall thickness of 0.5 mm, and a pinhole diameter of 0.8 mm. This very large pinhole was set up here with a two-stage dilution system: the carrier gas was set to 4 Nl/min (at 273 K) as for the first dilution and to 65 Nl/min in the second dilution stage [32,37,45]. An overall dilution of 500 was achieved.

Particles sampled from the flame were sent to a nano-DMA (TapCon 3/150 DMA system with a nominal size range of 2–100 nm equipped with a Faraday Cup Electrometer detector). In order to charge the particles to Fuchs' steady-state charge distribution [51] a Soft X-Ray Advanced Aerosol Neutralizer (TSI model 3088) was used. The PSDs obtained averaging over 3 scans were highly repeatable. The PSDs obtained by DMA were corrected for losses in the pinhole and the probe following the procedure reported in the literature [52–54]. DMA separates particles based on their mobility diameter so that the particle diameter could be retrieved from the correlation proposed by Singh et al. [55]. The uncertainty can derive from the evaluation of the wall losses and the coagulation of the small particles onto the large ones.

A 75×25×1 mm glass plate was horizontally inserted into the flame for 2 s to collect material from the flame at the highest equivalence ratio. The operation was repeated several times with a cooling cycle at room temperature of 10 s after each insertion. The procedure has been tested and validated before [7,56]. Spectroscopic techniques were used to characterize the sampled material similarly to what has been done for particles collected in flame fueled with other oxygenated fuels [33] and benzene [56].

Raman spectra were measured directly on the carbon samples deposited on a glass plate using a Horiba XploRA Raman microscope system (Horiba Jobin Yvon, Japan) equipped with a frequency-doubled Nd:YAG solid-state laser (λ=532nm) [33]. Raman spectra analysis provides information on the features of carbon network. FTIR and UV–visible (UV-Vis) spectroscopy were performed on the samples removed from the glass plate. FTIR spectra in the 3400–600 cm$^{-1}$ range were acquired in the transmittance mode using a Nicolet iS10 spectrophotometer. For FTIR analysis a sample preparation is needed; in particular, carbon particulate matter samples were mixed and ground in KBr pellets (0.2–0.3 wt%) [57]. FTIR gives information on the presence of oxygen within the carbon network. For the UV-Vis analysis carbon particulate matter samples were suspended in N-methyl-2-pyrrolidinone (NMP, with a concentration of 10 mg/L) and analyzed in a 1-cm path length quartz cuvette using an Agilent UV–vis 8453 spectrophotometer. The UV-Vis spectra were measured also on the soot fraction <20 nm obtained by filtration on an Anotop filter (Whatman) of a 100 mg/L total particulate suspension. UV-Vis analysis provides information on the mass absorption coefficients for the particles sampled indicating the level of aromaticity.

3. Numerical modeling

*3.1 Gas phase kinetic mechanism and soot model*

A detailed kinetic mechanism is used in this work, which includes the kinetic mechanism developed by D'Anna and coworkers [33,38,39,58] and the OME$_{1-3}$ kinetics taken from Sun et al. [29]. The complete mechanism consists of 141 species and 674 reactions, of which 41 species and 213 reactions are added to include the OME$_{1-3}$ oxidation kinetics [29].

The physico-chemical soot formation model employed in this work [38] has been combined with the CQMOM approach to track particulate evolution in [40] and successfully applied for simulating lightly- and highly-sooting flames in [40,41]. The present model distinguishes between different particle structures based on their state of aggregation, i.e. high molecular mass aromatic molecules (*molecules or large PAHs*), clusters of molecules (*clusters*) and agglomerates of particles (*aggregates*) [38]. This allows to follow not only the mass of the formed particles but also their hydrogen content and internal structure. Oxidation-induced fragmentation is also accounted for. Oxygen is considered the only species able to not react on the surface and diffuse towards the points of contact of the primary particles causing internal oxidation and subsequent particle fragmentation. PAH formation is modeled by the HACA and the resonantly stabilized free radical (RSFR) mechanism [38] and the molecular growth is described from benzene (A$_1$) up to pyrene (A$_4$). All the PAH compounds with a molecular weight larger than A$_4$ are not treated as individual species but considered as lumped species (*large PAHs*), whose evolution is described by the CQMOM. If the Van der Waals forces are strong enough to hold together these large molecules, *clusters* are formed. *Clusters* can grow via chemical pathways and can interact through coagulations. When larger clusters are formed, coagulation becomes an aggregation process, forming chain-like shaped soot particles (*aggregates*). Their reactions are described based on similarity with gas-phase kinetics following Arrhenius-rate laws.

*3.2 CQMOM model*

Large hydrocarbons and particulates not directly solved in the gas-phase kinetics are described by the evolution of the population balance equation (PBE) for the number density function (NDF) $f(\underline{\xi}; \underline{x}, t)$. Here, $\underline{x}$ is the space vector, $t$ is the time, and $\underline{\xi}$ is the internal coordinate vector defined as $\underline{\xi} = [\xi_{nc}, \xi_{H/C}, \xi_{stat}, \xi_{typ}]^T$, where $\xi_{nc}$ is the number of carbon atoms, with $\xi_{nc} \in [0, \infty)$, $\xi_{H/C}$ is the H/C ratio with $\xi_{H/C} \in [0,1]$, $\xi_{stat}$ represents the state of the particular entity with $\xi_{stat} \in A$, $A = \{stable, radical\}$, and $\xi_{typ}$ is the type of the entity with $\xi_{typ} \in B$, $B = \{large\ PAHs, clusters, agglomerates\}$. Note that the four coordinates are independent between each other and the two coordinates $\xi_{stat}$ and $\xi_{typ}$ have a discrete nature. Using the concept of the conditional density function allows writing the quadrivariate NDF, in which $\underline{x}$ and $t$ dependencies are dropped for convenience, as

$$f(\underline{\xi}) = f_{H/C}(\xi_{H/C}|\xi_{nc}, \xi_{stat}, \xi_{typ}) \cdot f_{nc}(\xi_{nc}|\xi_{stat}, \xi_{typ}) \cdot n(\xi_{stat}, \xi_{typ}), \qquad (1)$$

where $f_{H/C}$ represents the distribution of $\xi_{H/C}$ conditioned to a certain state ($\xi_{nc}, \xi_{stat}, \xi_{typ}$) and $f_{nc}$ represents the distribution of $\xi_{nc}$ conditioned on a state ($\xi_{stat}, \xi_{typ}$). The joint bi-variate distribution $n(\xi_{stat}, \xi_{typ})$ can assume only six different values, one for each possible combination $(u, v)$ of their discrete domain parameters. Therefore, evaluating $n(\xi_{stat}, \xi_{typ})$ at the six discrete points $(u, v)$ allows to reformulate the quadrivariate $f(\underline{\xi})$ as six bivariate NDFs $\Pi_{(u,v)}$

$$\Pi_{(u,v)}(\xi_{nc}, \xi_{H/C}) = f_{H/C}^{u,v}(\xi_{H/C}|\xi_{nc}) \cdot f_{nc}^{u,v}(\xi_{nc}) \cdot n_{u,v}. \qquad (2)$$

This is numerically convenient for the moment-based CQMOM approach used in this work. However, the six bivariate distributions $\Pi_{(u,v)}$ are not independent of each other, but they are strongly coupled by different source terms, e.g. the condensation of large PAHs on agglomerates, which modifies both NDFs. A complete description of the coupling approach is described in [40]. Similarly to [41], transport equations for the moment $m_{u,v}^{\frac{k_1}{z},k_2}$ of the six bivariate NDFs in Eq. (2) are solved

$$m_{u,v}^{\frac{k_1}{z},k_2} = \int_0^\infty \int_0^1 \xi_{nc}^{\frac{k_1}{z}} \xi_{H/C}^{k_2} \Pi_{(u,v)}(\xi_{nc}, \xi_{H/C}) \, d\xi_{H/C} d\xi_{nc}. \tag{3}$$

In the present study, fractional moments with $z = 3$ are used for the property $\xi_{nc}$. The moment inversion for fractional moments in the CQMOM context has been extended in [41] and is used here. When the moment inversion is completed, the weights $w$ and nodes $\xi$ are employed to determine the unclosed moment source terms for all the physical and chemical processes described in [38] and treated here in a lumped formulation. Two-way coupling between soot-phase and gas-phase is considered. The molecular diffusion of soot is neglected, as well as the thermophoresis which is known to have negligible effects in premixed flames [59]. Following Salenbauch et al. [41], the entropy maximization (EM) concept is employed in post-processing to evaluate the PSD, without prescribing the distribution shape. It is noteworthy that solving for the fractional moments gives the advantage to directly evaluate the PSD applying the EM concept. Indeed, using the relation between the equivalent-volume sphere diameter $d_p$ and the coordinate $\xi_{nc}$

$$d_p = \left(\frac{6W_c}{\pi \rho_s}\right)^{1/3} \xi_{nc}^{1/3} = L^{1/3} \xi_{nc}^{1/3}, \tag{4}$$

with $W_c$ the mass of a single carbon atom and $\rho_s$ the soot density, the transported fractional moments $m_{u,v}^{\frac{k_1}{z},0}$ can be directly transformed in diameter-based moments $\langle m_{u,v}^{k_1} \rangle$

$$\langle m_{u,v}^{k_1} \rangle = \int_0^\infty d_p^{k_1} f_{d_p}(d_p) dd_p = L^{k_1/3} \int_0^\infty \xi_{nc}^{k_1/3} f_{nc}(\xi_{nc}) d\xi_{nc} = L^{k_1/3} m_{u,v}^{\frac{k_1}{z},0}. \tag{5}$$

The diameter-based moments are then employed with the EM concept for evaluating the NDF at each location in the domain. Further details can be found in [40,41].

4. Results and discussion

Burner-stabilized ethylene/air and ethylene/OME$_3$/air flames with different equivalence ratios ϕ (see Table 1) are investigated both experimentally and numerically. One-dimensional simulations have been performed imposing temperature profiles measured in the experiments. The temperature in the current configuration is strongly influenced by the heat exchange between the flame and the burner. The chosen flame conditions exhibit negligible differences in temperature when OME$_3$ is added, similarly to other alternative fuels investigated before [6,7,10,32–36]. Finally, the same temperature profiles are used at each equivalence ratio for pure ethylene and OME$_3$ added flame. The CQMOM-based soot model is applied to solve the PBE for large PAHs, particle clusters and agglomerates. Two quadrature nodes for $\xi_{nc}$ and one quadrature node for $\xi_{H/C}$ conditioned on each $\xi_{nc}$ node for all of the six combinations (state, type) are used, which lead to an overall number of 36 moment transport equations. A variable soot density $\rho_s$ is considered between 1000 and 1800 kg/m$^3$ as an inverse function of the H/C ratio.

*4.1 Soot evolution*

In Fig. 1 the comparison between LIF and LII signals and the respective simulated volume fraction is only qualitative and reported for two equivalence ratios, 2.16 and 2.46. It is worth noting that the pure ethylene flame at an equivalence ratio of 2.16 is the first flame condition where LII has been detected above the noise level.

According to previous studies, LIF signals are here attributed to aromatic hydrocarbons in condensed-phase nanostructures that are not able to incandesce (see [2,37] and references therein). This attribution significantly affects the selection of adequate species from the modeling results for a comparison with the LIF signal. In previous works the LIF signal has been assumed to be directly related to the formation of condensed-phase nanostructures, i.e. particles with diameter $d_p \lessapprox$ 7 nm [41] while the detection of LII signal indicates the presence of larger particles [10]. Following [41], the simulated PSD obtained with CQMOM and EM, as described in Sec. 3.2, is split in a post-processing step to account for small particles with diameter $d_p < d_{p,slit}$ and larger particles $d_p > d_{p,slit}$, for comparison with LIF and LII signal, respectively. The value of the split diameter has been varied within the relevant range of 2 nm $\leq d_{p,slit} \leq$ 7 nm and the results are indicated by the error bars in Fig. 1. The CQMOM model is able to predict a substantially unchanged volume fraction of small particles (as from LIF) when OME$_3$ is added for both equivalence ratios shown in Fig. 1 (top).

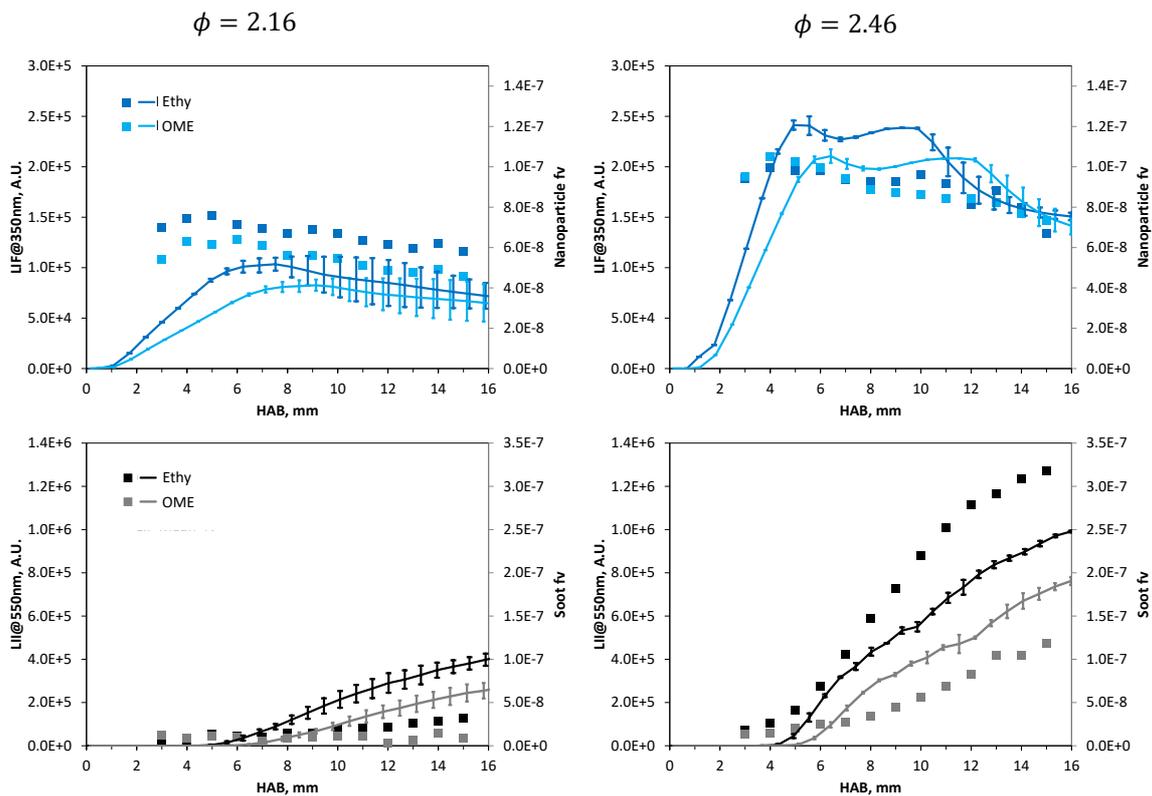

Figure 1 Comparison of the soot volume fraction predicted by the model with the experimental measured LIF signal (top) and LII signal (bottom) for $\phi = 2.16$ (left) and 2.46 (right) flames: ethylene/air (blue line and symbols: LIF; black line and symbols: LII), ethylene/OME$_3$/air (cyan line and symbols: LIF; grey line and symbols: LII). A separation is done for the particles calculated by CQMOM based on the PSD reconstruction with EM: for LIF, $d_p < d_{p,slit}$, for LII, (bottom) $d_p > d_{p,slit}$. Error bars indicate the sensitivity with respect to the value of $d_{p,slit}$ in the range 2 nm $\leq d_{p,slit} \leq$ 7 nm.

Furthermore, the simulations are able to predict the trend of increasing volume fraction of large particles (as from LII in Fig. 1) as the equivalence ratio increases as well as the reduction of large particles when OME$_3$ is added. However, the simulations predict a smaller reduction of large particles than what is experimentally observed.

The effects of OME$_3$ shown so far have been similarly observed in ethylene flames doped with other oxygenated compounds such as ethanol [35], dimethyl ether [36], butanols [10] and furans [33]. This suggests as for the other fuels that the main reason for the particle reduction has to be attributed to the modification of gas-phase precursors when OME$_3$ is added. Hence it is possible that the gas-phase model adopted here, developed mostly at low pressure environment [29], needs to be updated to predict gas-phase compounds.

Figure 2 shows the PSD measured at a height above the burner (HAB) of 15 mm for the four equivalence ratios investigated, with and without OME$_3$ addition. The measurements are compared with the predicted PSD from the EM procedure applied to CQMOM flame solutions. For soot particle aggregates, the simulated PSD are plotted versus the mobility diameter $d_m$, here equal to the collision diameter $d_c$ as in [60], $d_m \equiv d_c = d_p\, n_p^{1/D_f}$, where $n_p$ is the number of primary particles in an aggregate calculated on the base of the ratio between the mass of the aggregate and the mass of the primary particle. In this work, a fractal dimension $D_f$ equal to 1.8 and primary particle diameter $d_p$ equal to 15 nm have been assumed.

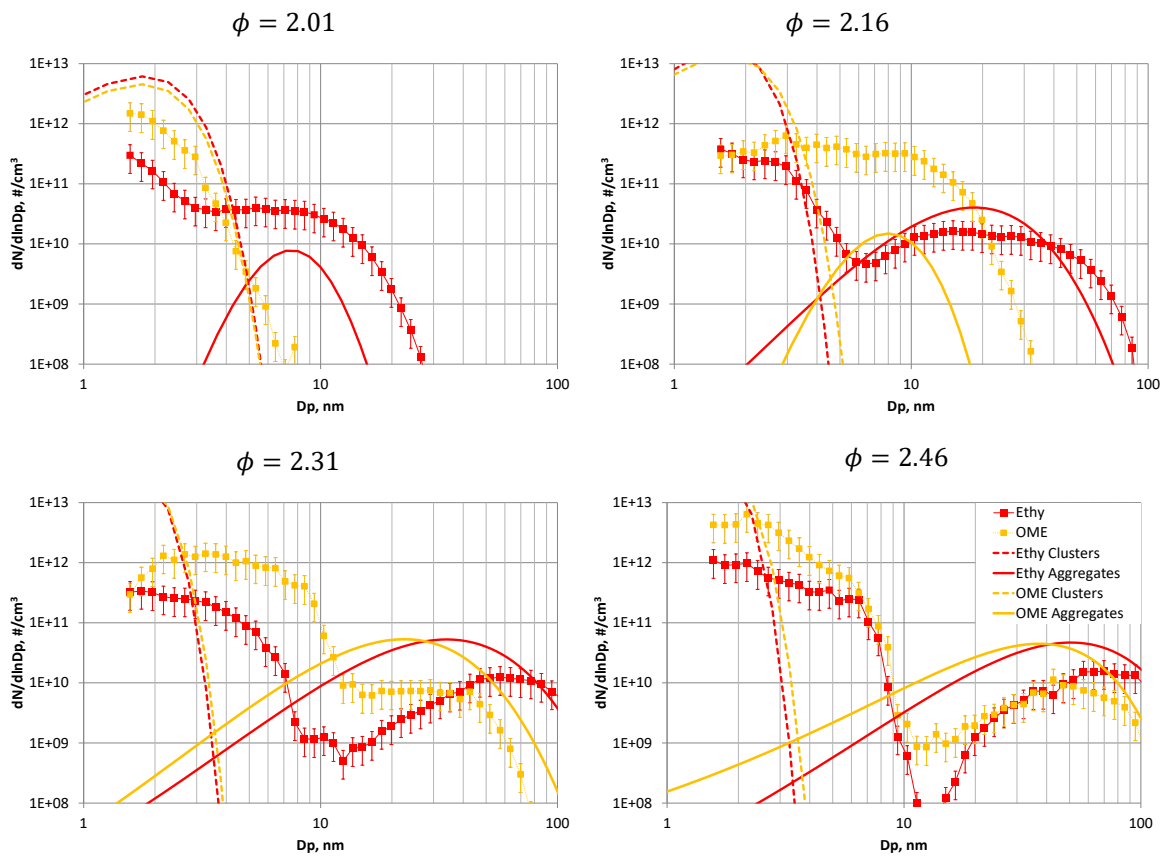

*Figure 2 Particle size distributions determined with EM on the CQMOM simulation results compared with the experimental measurements (SMPS) at HAB=15 mm for different equivalence ratio 2.01, 2.16, 2.31 and 2.46 flames: ethylene/air (red line and symbols), ethylene/OME$_3$/air (orange line and symbols). Dashed lines indicate the clusters and solid lines the aggregates.*

It can be observed that for slightly sooting conditions with $\phi = 2.01$ the OME$_3$ addition changes the shape of the PSD from bimodal to unimodal. A bimodal PSD remains evident, instead, for higher equivalence ratio conditions. An additional effect of the OME$_3$ blending is that a slightly higher number of small particles with $d_p < 5$ nm is formed, while larger particles with $d_p > 20$ nm are drastically reduced at all equivalence ratios.

The modeled PSDs indicate that all relevant effects of the $OME_3$ addition, described above, are qualitatively captured. In particular, the model is able to reproduce the trend of increasing size and the total amount of aggregates at increasing equivalence ratio and a substantially unchanged total number concentration of particles with a size smaller than 10 nm. However, the number concentration of the small particles is overpredicted at all equivalence ratios. It has been also reported above and in previous work [18, 24] that the SMPS measurements may be affected by significant losses of small particles.

*4.1.1 Modeling tests*

In order to further investigate the importance of the different kinetic pathways that lead gas-phase decomposition of $OME_3$ and their role on the particle formation, a numerical experiment has been performed. $OME_3$ has been numerically fed to the flame decomposed into smaller hydrocarbons to simulate a fast decomposition towards final products. The richest condition ($\phi$ =2.46) has been chosen for this study since the largest discrepancy with the experimental data in terms of soot reduction has been found (see Fig. 1). In particular, three possible cases have been analyzed.

$$OME_3 => CH_2O + CH_4 \qquad (1)$$
$$OME_3 (+O_2) => CO + H_2 + H_2O \qquad (2)$$
$$OME_3 (+O_2) => CO_2 + H_2O \qquad (3)$$

It is possible to see that the case 1 is a pure decomposition whilst the others are a decomposition with a partial (case 2) or complete oxidation (case 3). In all the decompositions the cold gas velocity has been kept the same by adjusting the nitrogen flow rate and, overall, the same moles of C, H, and O have been fed to the system. In fact, when $OME_3$ has been considered partially or fully oxidized the oxygen in the fed stream has been adjusted accordingly. For sake of clarity, in Table 2 the molar fractions of the fed streams in the decomposed tests are reported together with the base case. The temperature profile has been kept equal to the base case in order to evaluate only the chemical effects.

In Case 1, $OME_3$ has been decomposed in 4 molecules of $CH_2O$ and 1 molecule of $CH_4$; this is the simplest possible decomposition and it simulates the fast break of the C-C bonds. In Case 2, $OME_3$ has been broken and partially oxidized: all the carbon atoms coming from $OME_3$ are considered to be partially oxidized to CO, the hydrogen atoms coming from $OME_3$ are fed both as $H_2$ and $H_2O$. The amount of $H_2$ converted to $H_2O$ has been decided by leaving the $C_2H_4/O_2$=0.82 (i.e. $\phi$ for the ethylene equal to 2.46 – see Table 2). This decomposition simulates the capability of $OME_3$ to go towards partially oxidized products. Finally, in Case 3, $OME_3$ is considered to be fast enough to go towards the fully oxidized products and $OME_3$ is decomposed in $CO_2$ and $H_2O$. The selected pathways are here used to suggest whether considering a faster oxidation of the $OME_3$ is resulting in a further reduction of particles formed with respect to the current mechanism.

In Fig. 3 modeling results for the base case and the three decomposition cases are reported in comparison with experimental data: in the upper panel, the comparison with LIF and in the lower panel the comparison with LII are reported. It is possible to see that not all decomposition cases lead to an increase in particle reduction. In particular, when $OME_3$ is decomposed in smaller highly reactive hydrocarbons ($CH_2O$ and $CH_4$) the results do not significantly differ from the base case. This suggests that $OME_3$ decomposition - according to the kinetic scheme adopted here - is already fast enough to form small hydrocarbons. The slighter increase can be associated with the higher propensity to form radicals for molecular growth associated with $CH_2O$.

Looking at the second decomposition (the green line in Fig. 3), the reduction of both nanoparticles and soot particles with respect to the base case is significant. In this case, the presence of stable species (CO and $H_2$) is chemically slowing down the molecular growth process (radical less abundant).

Finally, when $OME_3$ is decomposed and fully oxidized, hence CO and $H_2O$ are added to the system, an increase in particle production is observed. The increase in particle formation has to be linked with the fact that the remaining ethylene and oxygen are burning in a much richer - although diluted by $CO_2$ and $H_2O$ – environment. The increase of the equivalence ratio overwhelms the dilution effect and the chemical effect of $CO_2$ and $H_2O$ that would lead to a reduction of particles. Overall the numerical tests suggest that the kinetic scheme used in this work is probably underestimating the capability of $OME_3$ to go towards partially oxidized products. It is possible that in mid-high temperature and oxygen-rich environment conditions - such as in the preheating zone of the investigated flame – a faster/direct formation of partially oxidized products could be favored. Future analysis and comparison with experimental data on gas-phase products could help to shed light on this point.

Table 2 Flame conditions for the decomposition cases. Inlet gas velocity @STP = 10 cm/s.

|  | Base case $OME_3$ | $CH_2O+CH_4$ | $CO+H_2+H_2O$ | $CO_2+H_2O$ |
|---|---|---|---|---|
| $C_2H_4$, mol frac | 0.1175 | 0.1175 | 0.1175 | 0.1175 |
| $O_2$, mol frac | 0.172 | 0.172 | 0.1433 | 0.1015 |
| $N_2$, mol frac | 0.6987 | 0.6517 | 0.6099 | 0.6517 |
| $OME_3$, mol frac | 0.0118 | - | - | - |
| $CH_4$, mol frac | - | 0.0118 | - | - |
| $CH_2O$, mol frac | - | 0.047 | - | - |
| CO, mol frac | - | - | 0.0588 | - |
| $H_2$, mol frac | - | - | 0.025 | - |
| $CO_2$, mol frac | - | - | - | 0.0588 |
| $H_2O$, mol frac | - | - | 0.0455 | 0.0705 |
| *C, mmol/s* | *2.97* | *2.97* | *2.97* | *2.97* |
| *H, mmol/s* | *6.18* | *6.18* | *6.18* | *6.18* |
| *O, mmol/s* | *3.96* | *3.96* | *3.96* | *3.96* |

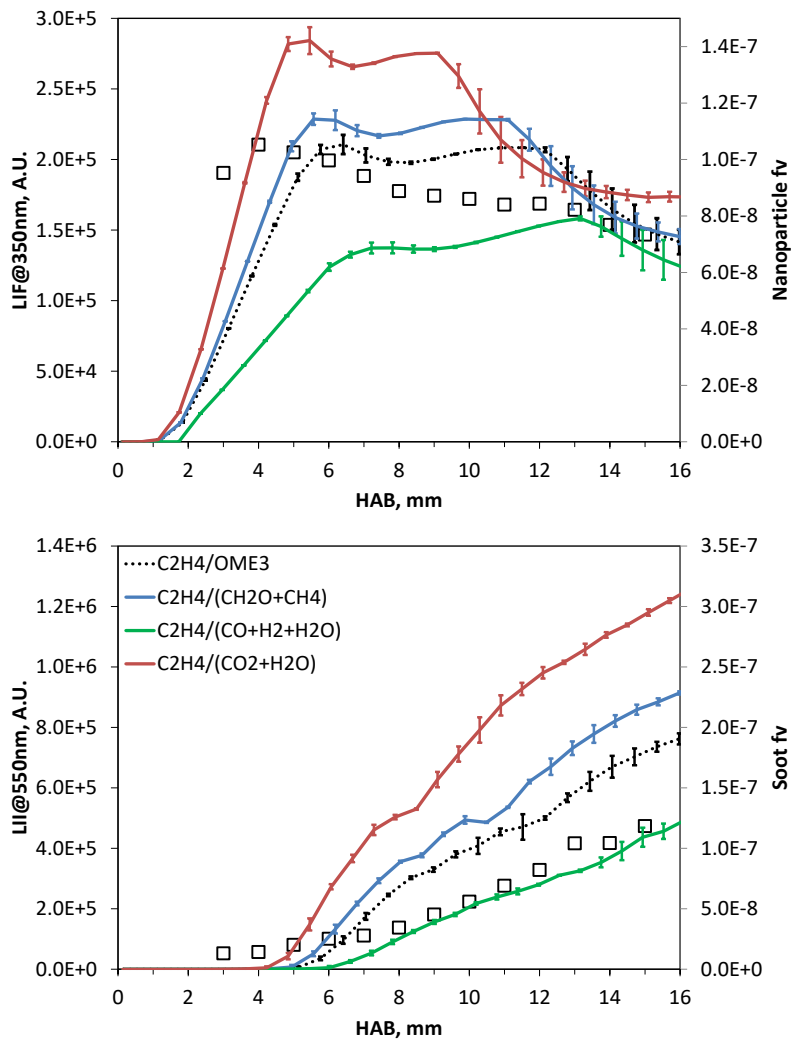

*Figure 3 Comparison of soot volume fractions predicted by the model with the experimental measured LIF signal (top) and LII signal (bottom) for $\phi = 2.46$ flames of ethylene/OME$_3$/air (empty symbols). The different coloured continuous lines represent modelling results for the different decomposition cases as reported in Table 2, dotted line represent modelling results for the base case. A separation is done for the particles calculated by CQMOM based on the PSD reconstruction with EM: for LIF, $d_p < d_{p,slit}$, for LII, (bottom) $d_p > d_{p,slit}$. Error bars indicate the sensitivity to the value of $d_{p,slit}$ in the range 2 nm $\leq d_{p,slit} \leq$ 7 nm.*

*4.2 Soot structure analysis*

A detailed spectroscopic analysis of the thermophoretically collected samples has been performed to verify the OME$_3$ effect on particulate properties. The analysis has been carried out on carbon particulate matter collected on a glass plate at 15 mm HAB and ϕ=2.46 in the pure ethylene and ethylene/OME$_3$ flames. UV–Vis spectroscopy has been performed on NMP suspensions of known concentration of carbon particulate matter removed from the glass plate. UV-Vis mass absorption coefficients are reported in Fig. 4 in the 250–900 nm range. The spectrum of particles sampled in the ethylene/OME$_3$ flame in comparison with that measured for the ethylene flame presents slightly lower mass absorption coefficients values indicative of a lower aromaticity. On the other hand, the UV-Vis spectrum values of the soot fraction <20 nm filtered from the total ethylene/OME$_3$ particulate appear more intense with respect to those measured for the pure ethylene flame, as shown in Fig. 5. In view of the similarity in terms of the shape of the bulk absorption spectra reported in Fig. 4, the lower absorption intensity for the ethylene/OME$_3$ flame can be justified by the larger abundance of the less light-absorbing particles <20 nm [61] in this flame, as also confirmed by PSD reported in Fig. 2.

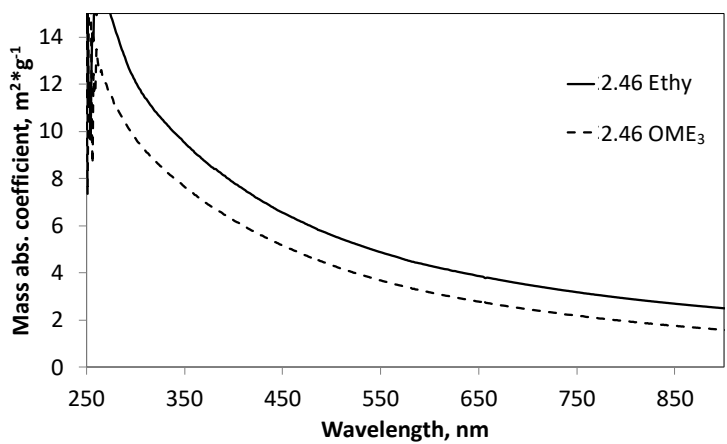

*Figure 4 UV-Vis mass absorption coefficient of particulate collected at 15 mm HAB in the ethylene/air (continuous line) and ethylene/OME₃/air flame (dashed line) at $\phi$ =2.46.*

The upper panel of Fig. 6 reports the first order region of the Raman spectra of particles sampled in ethylene and ethylene/OME$_3$ flames. The spectra present the typical features of disordered/amorphous carbons: a G peak at 1600 cm$^{-1}$ (vibration mode involved with the in-plane bond-stretching motion of pairs of C sp$^2$ bonds), and a D peak at 1350 cm$^{-1}$ linked with the disorder present in the carbon network [62]. The spectra have been normalized on the G peak to allow an immediate comparison of the spectral features. Regarding the peak position, a higher G peak position for the particulate collected in the OME$_3$/ethylene can be noted. It has to be reminded that, at whichever excitation energy, the G peak position takes a fixed value of 1580 cm$^{-1}$ in perfect and infinite graphite crystals [62], upshifted to 1600 cm$^{-1}$ for microcrystalline graphite due to the finite crystal size [62]. The G peak position in carbon materials can be upshifted from the 1600 cm$^{-1}$ band limit of purely graphitic carbon by the presence of shorter sp$^2$ C-C bonds, featuring olefinic bonds and/or small aromatic layers ([63,64]). Hence the higher G peak position of OME$_3$/ethylene soot (1606 cm$^{-1}$) in comparison to pure ethylene soot (1600 cm$^{-1}$) suggests a higher abundance of olefinic bonds (i.e. a lower abundance of sp$^2$ aromatic carbon) and/or smaller aromatic layer sizes.

The ratio of the D to the G peak intensity, I(D)/I(G), is the main Raman parameter used for quantifying order/disorder giving a quantitative measure of the size of the aromatic sp$^2$ phase organized in clusters. For highly disordered carbons, characterized by an aromatic cluster size smaller than 2 nm, the I(D)/I(G) ratio increases linearly with the crystal area following the equation proposed by Ferrari and Robertson [62].

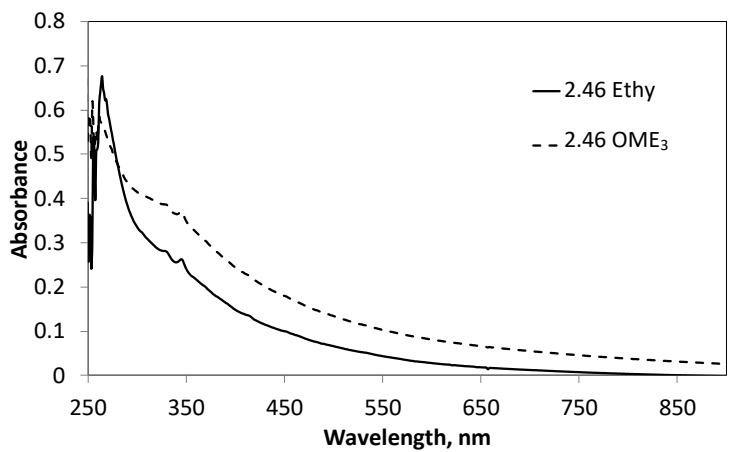

*Figure 5 UV–Vis absorption spectra of the filtered fraction (<20 nm) of 100 ppm of total particulate collected at 15 mm HAB in the ethylene/air (continuous line) and ethylene/OME3 /air flame (dashed line) at ϕ =2.46.*

As it can be observed in the upper part of Fig. 6 the spectrum of the OME$_3$/ethylene soot is characterized by a slightly lower intensity of the I(D)/I(G) ratio, that along with the slightly high position of the G peak demonstrates that the average aromatic island size within the carbon network in the ethylene/OME$_3$ particles is smaller. This suggests that the OME$_3$ addition slows soot aromatization and growth. These features can be associated with the general smaller sizes of particles collected in ethylene/OME flame with respect to pure ethylene flames. A shift of PSD towards small particles indicate a slowdown of the growth process that can be associated also with a lower level of aromatization.

Finally, to verify the effect of OME$_3$ on the composition of carbon particles, FTIR spectroscopy has been performed to identify the functional groups present on their surface [57]. The spectra have been measured using the same conditions, i.e. the same carbon concentration within KBr disk and disk thickness, to compare the results. More details on the sample preparation for FTIR measurements are available elsewhere [57].

The spectra in the FTIR spectral region 4000–2500 cm$^{-1}$ where OH stretching peaks (around 3500 cm$^{-1}$) and CH stretching peaks (3100-2800 cm$^{-1}$) occur, do not show significant differences and for this reason they are not reported. Thus, no significant differences in terms of hydrogen content were found, differently from our previous studies on 2,5-dimethylfuran- [7] and ethanol- [6] doped flames.

Infrared mass absorption coefficients of carbon samples collected in the ethylene and ethylene/OME$_3$ with ϕ = 2.46 are reported in the lower panel of Fig. 6 in the range 1900-900 cm$^{-1}$ mostly sensitive to the carbon skeleton and oxygen (C=O and C-O-C) functionalities and where some differences between the spectra are noticeable. These differences in the FTIR signal intensity are better shown in Fig. 6 also reporting the spectrum obtained as the difference between the infrared mass absorption coefficients of carbon particulate of ethylene/OME$_3$ flames and that of the pure ethylene flame.

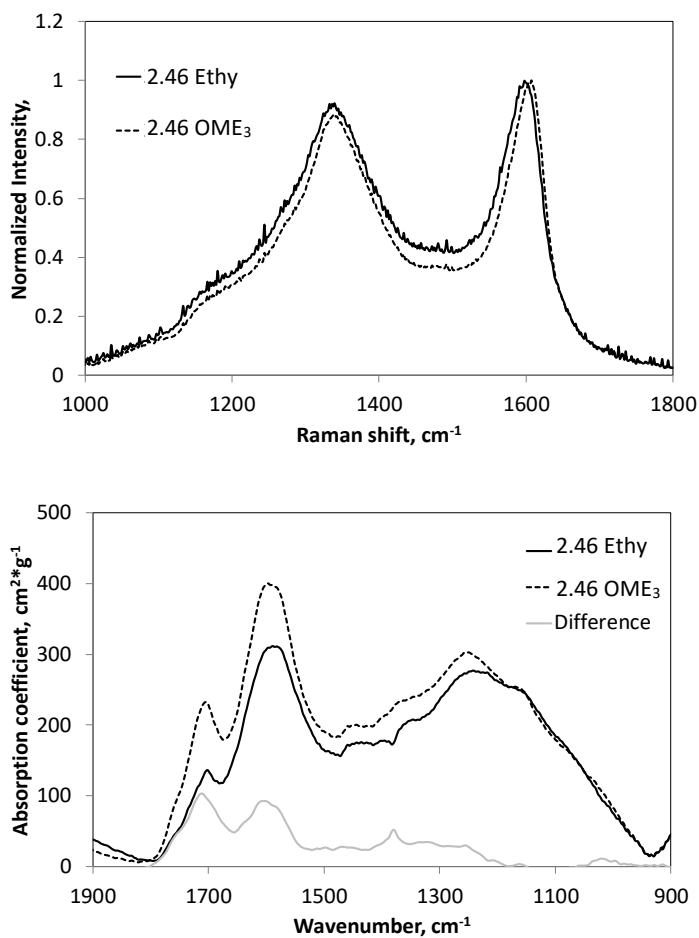

*Figure 6 Raman spectra (top) and infrared mass absorption coefficients (bottom) of soot particles sampled at 15 mm HAB and $\phi$ =2.46 in ethylene/air (continuous line) and ethylene/OME$_3$/air (dashed line) flames. The difference of the infrared mass absorption coefficient between ethylene/air and ethylene/OME$_3$/air flames is shown by the grey line (bottom).*

The higher intensity of the 1600 cm$^{-1}$ peak, due to the C = C stretching mode of polyaromatic systems, is clearly visible for particles collected in the OME$_3$ doped flame. The irregularity/dissymmetry of the aromatic moiety caused by whichever kind of ring substitution usually causes a strengthening of the 1600 cm$^{-1}$ peak. In particular, it was found that the increase in the dipole moment associated with ring vibrations in the presence of oxygen enhances the intensity of the C=C stretching peak [57,65]. To this regard soot sampled from the ethylene/OME$_3$ flame presents also a higher intensity of the 1720 cm$^{-1}$ absorption peak, due to the stretching of ketonic and/or esteric C=O groups [66].

Moreover, no significant differences were found in the 1300–1000cm$^{-1}$ range associated with other oxygen functionalities like ether type structures (C–O–C) as previously reported for butanols- [10] and dimethylfuran-[7] doped flames.

This is probably due to the different molecular structure of the fuel studied. In the case of OME$_3$ indeed it seems that oxygenated functionalities are not embedded inside the aromatic network of soot particles but are just located at the edge of the aromatic systems in form of C=O groups. It is not possible so far to conclude which are the pathways that lead to the presence of oxygen embedded into the particles. It is likely to be associated with the presence of oxy-PAH but on the exact nature of these PAH studies are ongoing. Numerical results and present gas phase scheme do not include the presence and/or an active role for oxy-PAH in particle formation. Their impact on the total amount of particles or even on the PSD is generally considerable less significative with respect to other combustion and kinetic parameters. However, further studies will have to deepen this topic also considering the significant role that the presence of oxygen can have in terms of after treatment systems and impact on human health [3,4].

## 5.     Conclusions

In this study, the effects of $OME_3$ addition on soot particle formation in burner-stabilized premixed ethylene flames have been investigated with experiments and numerical simulations based on the CQMOM approach. 20% of the total carbon has been substituted with $OME_3$ for four equivalence ratios, 2.01, 2.16, 2.31 and 2.46, while keeping constant the cold gas velocity. Experiments (LII and PSD) and numerical results indicate a reduction of the total amount and the size of soot particles in the $OME_3$ blended flame. An almost negligible effect has been observed on the small condensed-phase nanostructures, tracked by the LIF signal. Comparison with experimental data for pure ethylene and blended flames indicate a good predictivity of the model both in terms of soot volume fraction and PSD. However, at highly rich conditions the soot reduction due to $OME_3$ addition is underpredicted. This behavior has been hypothesized to be linked with the decomposition/oxidation of $OME_3$ in the early stage of combustion and subsequent formation of partially oxidized by-products and radicals. The gas-phase kinetic scheme for $OME_3$ taken from the literature was originally developed for different combustion conditions. To investigate the influence of the gas-phase reaction pathways on particle formation, numerical tests have been performed feeding, in place of $OME_3$, its decomposed or oxidized products to identify possible chemical pathways that lead to a reduction of particle growth. It has been observed that a faster formation of partially oxidized products in the gas-phase kinetics may favor a further soot reduction with respect to the actual scheme; however, further research is needed to investigate the gas-phase pathways.

Finally, the effects of $OME_3$ addition on the chemical features of the particles, i.e. aromaticity and composition, have been analyzed in the highest equivalence ratio condition. UV-Visible and Raman spectroscopy analysis suggest a slightly lower degree of aromatization in $OME_3$-doped flames, probably due to the higher concentration of particles with a size lower than 20 nm. A higher presence of C=O functionalities was found analyzing the FTIR spectra of the particle samples, while no significant differences were observed in C–O–C, OH, and CH functionalities.

**Acknowledgments**

The authors gratefully acknowledge the funding by the German Federal Ministry of Education and Research (BMBF) within the NAMOSYN Project (project number 03SF0566R0) and by the Clean Sky 2 Joint Undertaking under the European Union's Horizon 2020 research and innovation programme under the ESTiMatE project, grant agreement No. 821418.